Development and Analysis of P2SCP: A Paradigm for Penetration Testing of Systems that Cannot be Subjected to the Risk of Penetration Testing

Jeremy Straub
Institute for Cyber Security Education and Research
North Dakota State University
1320 Albrecht Blvd., Room 258
Fargo, ND 58108
+1 (701) 231-8196 (phone)
+1 (701) 231-8255 (fax)
jeremy.straub@ndsu.edu

**Abstract**

Penetration testing increases the security of systems through tasking testers to 'think like the adversary' and attempt to find the ways that an attacker would break into the system. For many systems, this can be conducted in a safe and controlled way; however, some systems are so critical to human life and safety that the risk of their failure or disablement due to active penetration testing cannot be assumed. These systems are also critical to evaluate the security of, to prevent attackers from disabling them or causing their maloperation; however, this must be done in a manner that doesn't risk the very malady that testing seeks to avoid through the testing process itself. This paper presents P2SCP, a paradigm for penetration testing of systems that cannot be subjected to the risk of penetration testing. It discusses how data collection, the creation of digital twins and cousins and evaluative analysis can be utilized to conduct virtual penetration tests on critical infrastructure systems. This proposed paradigm is analyzed through the use of several case studies.

**Keywords:** penetration testing, cybersecurity, critical infrastructure, risk, security assessment

**1. Introduction**

Critical infrastructure systems such as those that operate water supplies, power generation and distribution and natural gas are essential to the lives and safety of billions of people worldwide. These systems must work reliably and, in many cases, must work continuously. A key aspect of these operations is the cybersecurity of the systems.

Penetration testing is used to evaluate and enhance systems' security. Testers are tasked to 'think like the adversary' to attempt to break in to, disable or damage the system being tested to identify prospective vulnerabilities. By identifying them, they can be corrected before they are leveraged by a nefarious party as part of an attack.

Penetration testing, though, is not without risks – and the risks are inherently assumed by the operators of (and, indirectly, the users of and those relying on) the system under test. In some cases, though, the risk of system disablement is too great to be chanced. For example, if a penetration test was successful in disabling a power distribution facility in winter, residents of the area served by that facility might face the risk of freezing to death. While controls can be imposed to attempt to control the risks that may eventuate, the goal of penetration testing is to identify the unknown unknown. This, inherently, carries with it the risk of something unexpected happening, as testing is being conducted.

This paper proposes and evaluates a paradigm for penetration testing of systems that cannot be subjected to the risk of penetration testing (P2SCP) as a partial solution to the challenge of assessing the security of systems that cannot be subjected to traditional penetration testing. It continues, in Section 2, with a discussion of relevant prior work. Then, in Section 3, the paradigm is described in detail. Following this, in Section 4, use scenarios for the paradigm are discussed. Next, in Section 5, sources of input data for the paradigm are discussed. In Section 6, the operations of the paradigm are discussed. After this, in Section 7, the outputs of the paradigm and their implications are presented. Finally, the paper concludes, in Section 8, and discusses several potential areas of needed future work.

## 2. Background

This section presents prior work in several areas that provide a foundation for the current work. First, prior work regarding penetration testing is presented. Next, previous penetration testing automation is discussed. Finally, digital twins are reviewed.

### 2.1. Penetration Testing of Critical Infrastructure Systems

Significant amounts of focus have been devoted to securing and evaluating the security of critical infrastructure systems. A number of test labs and platforms have been developed [1,2] or proposed [3]. Ficco, Choras and Kozik [4], in particular, presented a concept for a "hybrid and distributed simulation platform" which combined simulation and actual hardware systems. Turpe and Eichler [5], on the other hand, proposed performing penetration testing on live systems while taking precautions and using methods to mitigate risks, such as restricting tests to only a "selection of test cases and techniques" that would be performed on isolated subsystems.

Several examples of simulation for critical infrastructure systems exist, such as implementing a "virtual plant environment" [6] and a security simulation system [7], based on a threat model and assessment activities. A technique called "Stackelberg planning" [8] was proposed which utilizes a model and pits defenders against simulated attackers. Another paper [9] suggested that penetration testing should be standardized using a proposed framework. Some, such as Li, Yan and Naili [10], have suggested artificial intelligence and deep reinforcement learning, be used to identify "optimal attack path[s] against system stability".

Brown, Saha and Jha [11,12] contributed the concept of assessing the impact of an attack on the entire system or network, instead of focusing solely on attack success, and Lee, et al. [11,13] suggested incorporated fuzzing to assess areas such as software defined networks.

### 2.2. Penetration Testing Automation

Automation offers the ability to help with a key area of critical infrastructure security: identifying previously unknown system vulnerabilities so that they can be remediated. Systems are, of course, likely to be vulnerable to unforeseen attacks [14] and attack types. As systems increase in complexity, the potential for complex exploits to be developed by attackers, which exploit multiple small, overlooked configuration issues or vulnerabilities increases. Thus, many critical infrastructure systems will benefit from a mechanism to automatically search for vulnerabilities [15] so that they can be secured.

Cyberattacks are becoming increasingly automated and rapid as are their countermeasures [16]. The very tools that enhance the speed and efficacy of attacks can also be leveraged to automate penetration testing to attempt to find issues and correct them before an adversary does. Reinforcement learning [17], for example, has been demonstrated for just this purpose. Artificial intelligence has also been used for automating vulnerability detection [18].

Attack tools are not the only area benefiting from automation-enhancement. Intrusion detection and prevention systems, which protect networks and detect attacks [19], have also been enhanced by artificial intelligence and data mining technologies. Artificial intelligence can also be used to identify unnoticed past attacks [19] and its use has been demonstrated for protecting industrial control systems [20].

Problematically, though, many tools may risk causing the very impact on the system being tested that the tests seek to protect against and avoid [21]. Thus, automated testing may not be prudent directly against live critical infrastructure systems.

### 2.3. Digital Twins

According to Wright and Davidson [22] digital twins are "essentially a model and some data"; however, their name invokes more and a digital twin should include "a model", "an evolving set of data" and "a means of dynamically updating or adjusting the model in accordance with the data". However, there is some disagreement over what is and isn't a digital twin since, as Cimino, Negri and Fumagalli [23] note "agreement over [their] features and scopes has not been reached". They have found use in a variety of areas including "smart manufacturing" [24], the "design and operat[ions] of cyber-physical intelligent systems" [25], engineering advancements [25], anomaly and fault detection [26], traffic management [26] and expediting analysis [26].

Digital twins trace their lineage back to NASA, with the Apollo program first using the "'twin' concept" [27] and the work of Grieves and Vickers introducing the concept for life-cycle analysis [28]. Hernandez and Hernandez [29] were the first to use "the 'digital twin' terminology", according to Lui, et al. [27].

Digital twins are used for safety analysis [23], prediction [30], risk and cost reduction [31], monitoring [23,30], optimization [23,30], simulation [30], maintenance [23], enhancing decision making [31], management [23] and increasing "service offerings", reliability, security, resilience and efficiency [31]. Thus, while their precise definition may still be elusive, their benefits are notable throughout a number of application areas in numerous industries.

### 3. Paradigm Description

The P2SCP paradigm is based on the functional modeling of cyber-physical or software systems. The idea of the functional model is to capture the functionality of every device in a system and the boundaries of the system to be able to anticipate how devices will function and interact under a number of circumstances. Figure 1 demonstrates this conceptually. It shows how a variety of types of inputs (such as network diagrams, system information and network scans) serve as inputs to three different types of models. These models can then be used to produce a number of different types of output products (such as Cyber Kill Chain analysis, supply chain analysis and attack detection indicators). These inputs will be discussed in Section 5, the outputs will be discussed in Section 4 and the model types will be discussed in Section 6.

The functional model is created with a small number of parts: containers, links and generic rules. These are part of the simulation system, which is used to implement the proposed paradigm. Containers are the objects in the network, such as computers, servers, routers and functional technology systems. They are defined by a collection of property values that store information about them and their state. Links model the physical interconnections of network objects. Finally, generic rules match links based on the properties of the containers that they connect. Generic rules can be used to model attacks and can change the state of the containers that they connect, if they are successfully run. A model constructed of links and containers is shown in Figure 2 and the definition of generic rules in shown in Figure 3.

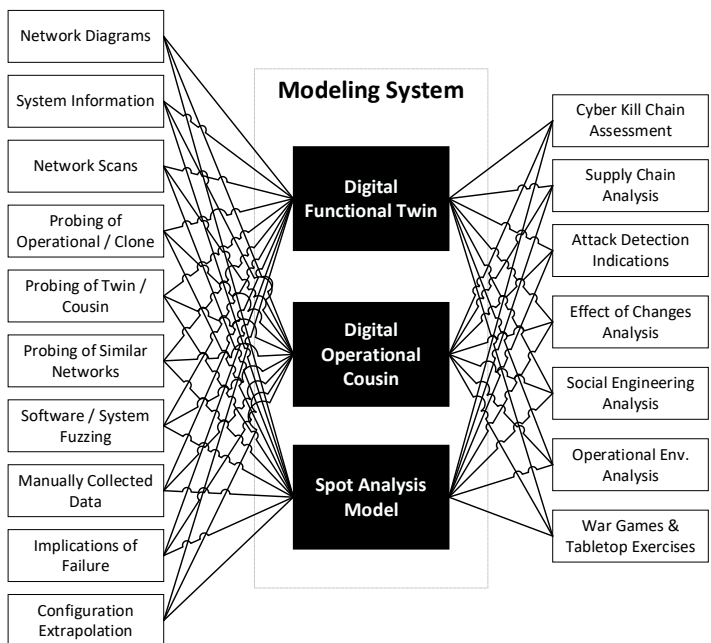

Figure 1. Modeling system inputs, model types and outputs.

The implementation of the functional model starts by providing a depiction of the system as it currently is (or as it was at the point in time that data was collected, if changes have been subsequently made to it). The model is updated as rules are triggered and change the properties of the containers within the network. Changed properties may cause other generic rules to now match a link and its associated containers, allowing additional rules to be run. In some cases, it may make sense for rules to be able to be run multiple times – in others, rules may be best implemented to be allowed to run only once.

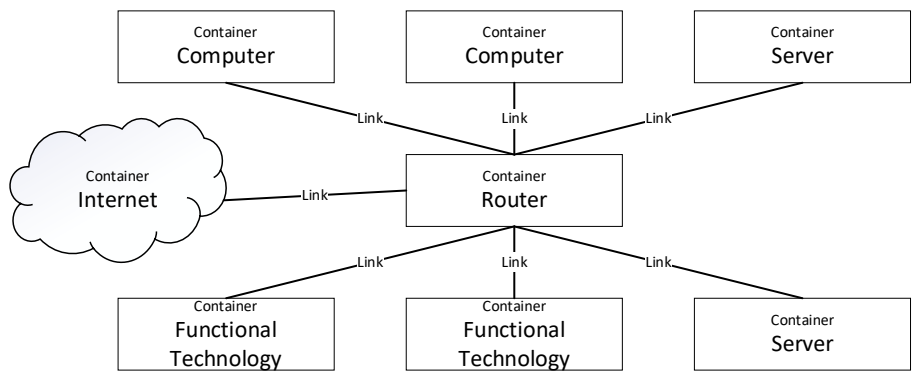

Figure 2. Implementation of a model as links and containers.

Depending on how the system is being used, rules can be triggered automatically based on their match conditions being satisfied (i.e., the linked containers have the required properties in the required states). Under this approach, all possible rules would run and operations would continue automatically until there were no further rules left to run. Alternately, the system can be utilized to assess and project the results of the actions of human attackers and rules can be selected to be triggered by these humans as part of a role playing (or war) game or tabletop exercise.

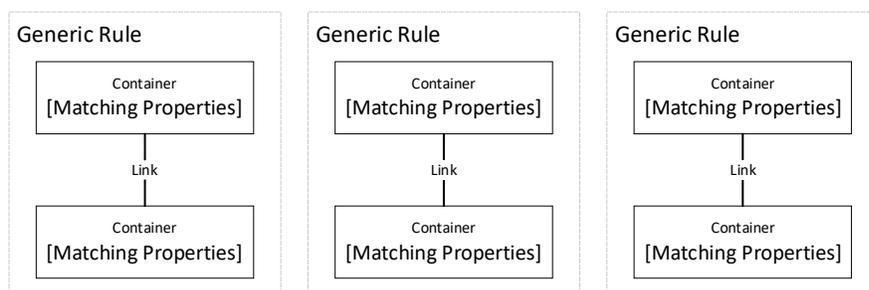

Figure 3. Implementation of attacks as generic rules.

Depending on the nature of the assessment being conducted, models can be defined to depict an entire system in as accurate a manner as possible. Alternately a model could be made of a system that is functionally similar to the system of interest. This model may be abstracted to remove details and redundancy, to facilitate a more generalized assessment or to safeguard key system information.

Within the scope of the modeled system (or modeled areas of a system, if only a part of the system is modeled), the functional model can be used for a number of purposes. These are discussed in Section 4.

## 4. Use Scenarios

This section discusses potential use scenarios for the proposed functional modeling system. These include Cyber Kill Chain assessment, supply chain assessment, attack detection, change effect determinization, social engineering impact determination, broader context evaluation and war game / tabletop exercise use.

### *4.1. Cyber Kill Chain Assessment*

Lockheed Martin introduced the Cyber Kill Chain as a mechanism to identify and combat pathways of cyberattack and the MITRE Corporation introduced a similar concept with their ATT&CK model. Both frameworks present a series of steps that an attacker must complete to conduct a successful cyberattack and access a system (with the MITRE ATT&CK model providing more detail on the specific techniques and tools that can be used to do so [32]). The Cyber Kill Chain model, for example, has seven phases: "reconnaissance", "weaponization", "delivery", "exploitation", "installation", "command and control" and "actions on objectives" [32,33]. While a simple attack might be modeled as a single kill chain with one instance of each of these steps, attacks against systems of any size inherently require multiple steps, which can be conceptualized as multiple iterative (or, in some cases, nested) kill chains.

The proposed paradigm and modeling system is specifically designed to facilitate modeling and threat assessment using complex kill chains. The IT system being evaluated is modeled as a collection of containers (with applicable properties) and links. Generic rules provide the mechanism for identifying methods of delivery, exploitation, installation, command and control and actions on objectives and can be applied to any applicable link.

Figures 4 and 5 illustrate this, conceptually. Figure 4 shows a basic IT system comprised of two computers, two switches and a router. The computers are physically connected to the switches and a router interconnects the two switches. Logically, this can be reduced to the model shown in Figure 5, as the switches are – in this instance – basic equipment that are not capable of performing filtering or supporting vLANs. Also, in this limited model, an attack to cause the switch to broadcast all traffic (as a way of an attacker getting access to it) would not be needed, given that the switches serve as two-device interconnections. Thus, while switches may need to be represented in some logical models, this model can be simplified to discount them, facilitating explanation simplicity and producing the model shown in Figure 5.

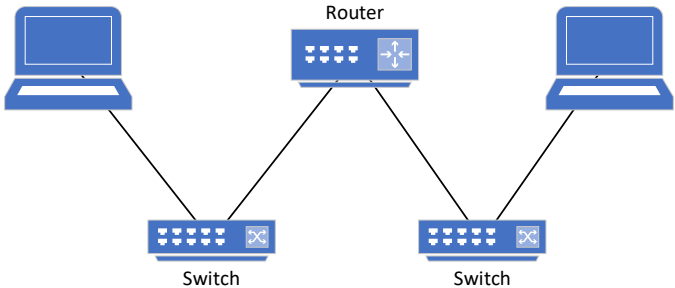

Figure 4. Example IT system.

If an attacker were going to try to use the left computer to attack the right computer, the efficacy of different prospective attacks could be evaluated using this model. Two links would need to be traversed. The attacker would need to traverse the link between the left computer and the router and the link between the router and right computer.

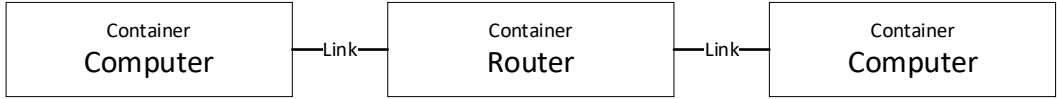

Figure 5. Model of example IT system.

In the most basic circumstance, where the router is configured to forward all traffic between the two computers, the first link is easily traversed, simply by sending network traffic. This would be represented by link properties showing the configuration and a basic data movement process to traverse the link. The second link has a data movement consideration as well as a need for a mechanism to access the right computer, in order to fully traverse it. Thus, unless the right computer was completely unsecured, an attack would need to be implemented to traverse the second link.

In a more complex case, where the router would not allow the traffic to pass under its default configuration, multiple attacks may be needed. First, the left computer would need to get access to the router, with an attack of some sort, and then either change the router's configuration to forward the requisite network traffic or use the router to attack the right computer directly. Again, an attack would be needed against the right computer. Thus, there might be three areas of attack need: accessing the router, privilege escalation or other attack on the router itself and an attack against the right computer (either from or via the reconfigured router).

To determine whether this can be successfully conducted or not, more information about the computers and router would be needed. Figure 6 presents some relevant properties for the left computer and the router. For example, it shows that the left computer has Metasploit installed and that the router has two vulnerabilities; however, at present, neither administrative or user level access is available to the router.

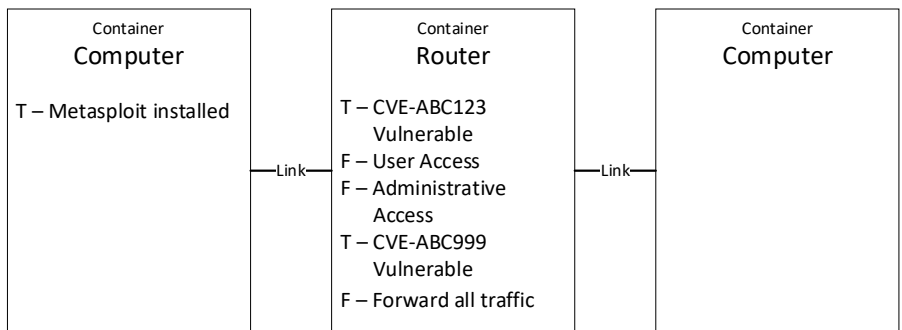

Figure 6. Model of example system with properties.

Three attacks that can be utilized are presented in Figure 7. Each is defined in terms of pre-condition and post-condition properties that linked containers must have in a particular state in order for the attack to be conducted.

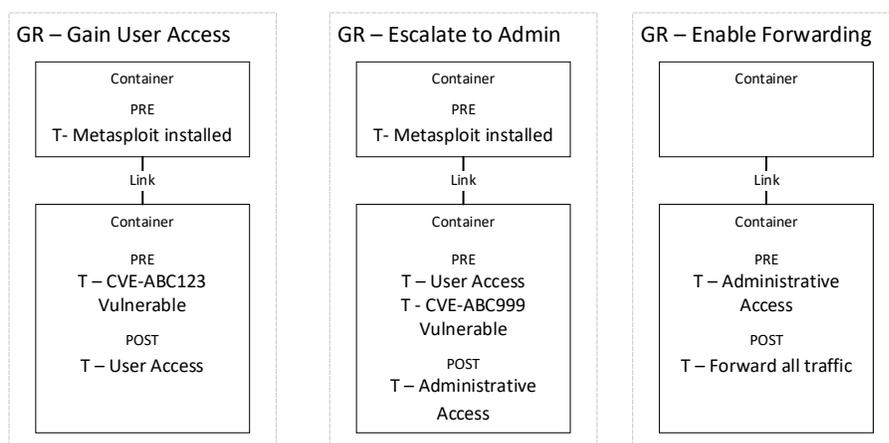

Figure 7. Example generic rules with pre and post properties.

Using these generic rule attacks, an attacker could first utilize the left computer to attack the router and gain user-level access (using the left-most generic rule attack which is conducted via Metasploit). Then, again using Metasploit, a second attack could be conducted – now that user-level access is available to escalate from user-level access to administrative level access. Finally, the right-most generic rule can be utilized (which is not really an attack, per se, but a configuration change) to enable forwarding. These three rules, when executed together and in the proper order, provide the left computer with access to attack the right computer.

All elements of a cyber kill chain (or MITRE ATT&CK)-based attack can be modeled in this way. Larger and more complex systems will inherently necessitate more complex attacks; however, the same concepts of modeling apply. The system is inherently scalable. It can also be used to model conceptual attacks by abstracting the system configuration and rules to a conceptual level.

Once the model is created, in addition to testing specific scenarios, numerous runs can be conducted across all possible attack paths and the frequency of individual nodes in those attack paths can be determined. This can be used to identify the most impactful places to enhance security in a computing system. Additionally, proposed security enhancements can be simulated to see what impact they have on the total number of possible attack paths and to see what nodes then become the most frequent attack path members, once a particular node is (or set of nodes are) better secured.

*4.2. Supply Chain*

Supply chain vulnerabilities and exploits are a critical area of consideration for critical infrastructure systems. They can come from a variety of different sources.

One source is adversary activities. An adversary could intentionally introduce a vulnerability through the substitution of a component for another or the substitution of an intentionally different or defective component for the correct one. Alternately, an adversary could cause a device to be incorrectly manufactured with added, removed or improperly connected components. These can be effected through intelligence activities, utilizing an adversary's direct control over a supplier or through the use of social engineering to effect component or device suppliers.

Adversary intervention may not necessarily be required to have an exploitable supply chain vulnerability, however.  An adversary could become aware of an issue that they had no involvement with the creation of, such as a latent defect in a part, a design flaw, an assembly flaw or an inadvertent and undocumented (or documented but unapproved) product change.  Adversaries could also become aware of nefarious supplier activities or intentions and make assumptions as to what these may translate to in terms of product differences (such as the potential impact of cost-cutting measures).

The proposed paradigm is designed to be able to consider supply chain vulnerabilities and their exploitation.  Notably, these can be considered as part of the aforementioned kill chain analysis and in conjunction with other types of analysis described herein.

To model supply chain dependencies, containers have been designed to be able to be nested within each other.  Figure 8 shows an example of components (in this case a number of arbitrary chips) that are part of a computer and router.  For simplicity, only a small number of chips are depicted and other considerations, like assembly and design, are not included – though they could be added for an actual analysis process.

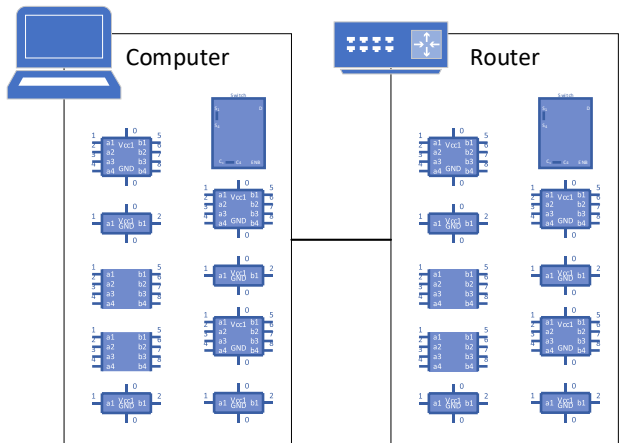

Figure 8. Example of supply chain comprised of computer and router components.

To model this, as shown in Figure 9, a number of containers are created and nested within the computer and router device containers.  Again, in this case, a number of chips are utilized.  These nested containers can have links between them, as well as – potentially – with other containers outside of the device.

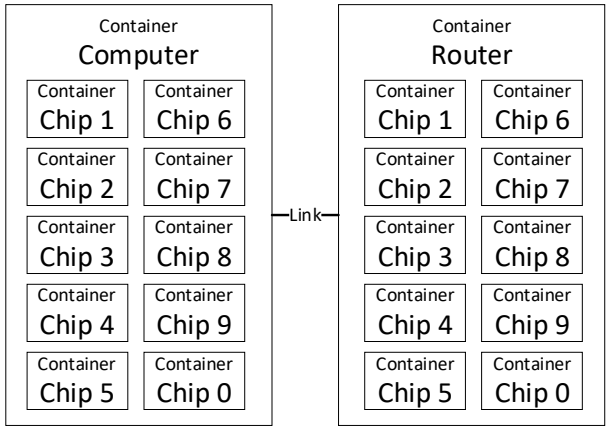

Figure 9. Model of supply chain comprised of computer and router components.

Just as generic rules were utilized to model attacks between device containers, they can also be used to model attacks and interactions between devices and their components (inclusive of design and assembly, in addition to physical hardware). Figure 10 shows three types of generic rules that can be used with components.

The left most generic rule example changes a property of one (parent or child) based on the property of the other. This type of rule can be set to be automatic (i.e., the change to the parent / child implicitly changes the other's property immediately) or to be triggered by a specific action that has been taken.

The middle generic rule example shows how a rule could be created that defines interacitons between multiple children and their parent. Again, these interactions could be automatic or triggered.

Finally, the right-most generic rule example shows how a rule could be run on two containers within a parent (potentially a parent with specific properties) container.

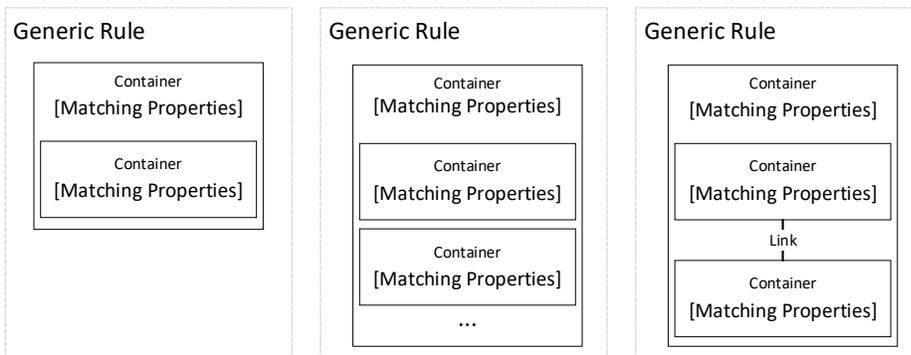

Figure 10. Example of supply chain generic rules.

In each instance, aside from the relationships between the nodes that they act on, each generic rule operates similar to the ones described in the previous section.

*4.3. Attack Detection*

The development of a model – including basic kill path models and models that include supply chain considerations – can have a number of uses, beyond simply identifying potential attack paths for defenders to inhibit (or for attackers to utilize).  One potential additional use of these models is for identifying signs of attack detection and facilitating attack type identification.  The goal of this is to be able to rapidly detect and identify an attack on a computing system based upon prior simulation of the results of attacks.

This simulation will identify areas to monitor for indications that a particular attack is ongoing or has been successful.  Then, by assessing the status of the network after the simulated attack, a signature for the attack's results can be identified.  When an actual attack is detected, these same monitoring locations – on the operational network – can have their status assessed.  Then, the attack can be identified by matching the actual symptoms to close matches from the simulation, which would be likely candidates for the type of attack that has been detected.

The simulation process is quite straight forward.  The network is run with a variety of simulated attacks being conducted.  For each attack, the results of all containers' properties are stored and differences from the starting configuration are identified.  Symptoms and combinations of symptoms that are unique (individually or collectively) become a signature for the attack type.

The simulation should be run and property data collected with attacks completing successfully, attacks that are not successful and stopped for data collection at different points in the attack.  This will facilitate the identification of attack attempts and failed attempts, in addition to completed attacks that have been successful.

### *4.4. Effect of Changes*

Another prospective use of the proposed paradigm is to evaluate the effect of potential changes on a computing system.  One potential use, as was briefly mentioned previously, is to assess the impact of security enhancements.  However, the proposed paradigm can be used to assess the security implications of any prospective system changes that can be incorporated into the model.  It can also be utilized to compare two prospective sets of changes to evaluate whether one produces a preferable outcome, as compared to the other.

The approach to doing this is also relatively simple.  First, metrics of evaluation need to be defined that the system can provide details regarding.  For example, the number of kill chain pathways identified could be a key metric.  Then, the system is run with the model of the computing system that is being assessed in its current state.  Data is collected regarding the identified key metrics to provide a baseline for the purposes of comparison.

Following this, the next step is to make all or part of the proposed modifications to the computing system and to collect metric data again, for comparison.  By implementing the modifications in partial groups and collecting metric data in between these implementation groups, the impact of each part of the changes can be independently assessed.  Thus, areas of particular benefit or negative impact can be further investigated for potential replication or mitigation purposes, respectively.

In addition to assessing changes based on identified metrics, the proposed paradigm can be used to prepare for securing and operating the changed system.  It can be used to see if new vulnerabilities have been introduced or existing vulnerabilities have been mitigated.  The change model can also be used to conduct node-involvement-in-attack-path analysis to see if nodes' involvement frequency changes.  This can be used to identify the best areas of the computing system to focus on the security and monitoring of after the proposed changes have been made (and to identify whether these areas of criticality will change due to the proposed changes).

*4.5. Social Engineering*

Social engineering threats can be particularly difficult to assess due to the fact that, in many environments, they can eventuate as an effect to virtually any part of a critical infrastructure system at any time. Acknowledging their ubiquity of potential impact helps demonstrate the importance of monitoring for them, raising awareness amongst employees about them and deploying policies, procedures and general-purpose technical safeguards to mitigate their occurrence and impact. It does little, however, to identify where they may be most likely to be targeted or what their impact would be if they occur.

The proposed paradigm presents a solution to this. Instead of attempting to identify every possible social engineering attack (of which there are a practically unlimited number of), it can be utilized to identify areas of an attack path that cannot be completed using technical means or areas where an attack path can be shortened through human action. Given their comparatively high cost and potentially higher risk, these are the areas that social engineering attacks are most likely to be utilized.

Like was discussed previously, in regards to technology-based attacks, the proposed paradigm can identify indications of these types of attacks being used currently and of prior successful and unsuccessful attacks of this type. It can also be used to help project social engineering attacks' impact for risk analysis, planning, decision-making and countermeasure implementation purposes.

The process used to do this is simple: attack paths are generated and an algorithm is utilized to scan to identify node connections that would allow other attack paths to be completed. An algorithm is, then, also run to look for areas where attack paths can be reduced in length through human intervention. These two sets of result areas are potential areas of social engineering use. Finally, applicable attacks can be manually added to the system in these areas and additional simulation runs can be performed to identify the impact of social engineering on nodes' attack path involvement levels, the overall level of attack paths generated, potential attack time benefits to attackers and to identify indications and warnings of these social engineering attacks being underway or having been conducted (successfully or unsuccessfully) in the past.

*4.6. Evaluation in the Context of Broader Operational Environment*

Conti and Raymond [34] define the "operational environment" as the "conditions, circumstances, and influences that affect military operations". Working from this definition, they explain how the U.S. Army's PMESII-PT framework can be used to understand the operational environment of cyberspace operations. This framework considers the "political", "military", "economic", "social", "information", "infrastructure", "physical environment" and "time" factors [34]. All of these are integral to cyberspace operations and the analysis of all of them can be aided by the proposed paradigm.

Each of these factors can prospectively serve as an input to the proposed paradigm's model and model outputs can, in many cases, aid analysis of each factor. Thus, in most cases, the operational environment context both serves as a foundation for and is impacted by the modeled critical infrastructure system.

The first four factors (political, military, economic and social) are logical inputs to the system. Conti and Raymond [34] discuss two principal areas of political consideration. The first is online political activities, such as voting and campaigning. The second is the influence of politics on the online environment through regulations, standards and other mechanisms. The military factor, they indicate, primarily relates to the size and capabilities of warfighting forces (both online and offline). The economic factor mirrors the political factor, including online economic activities (such as banking and sales) and the influence of economic activities (such as policies) on the online environment. Finally, the social factor also has similar considerations – such as the number of people with internet access, online cultures and the impact of social factors on online regulation.

The information factor includes both news content and information systems (inclusive of those providing commercial and military capabilities). It also includes public sentiment considerations and mechanisms of media control. The infrastructure factor focuses on utilities including those that supply power and interconnectivity for and satisfy other operational needs of information systems. The physical environment includes both geographic locations related to information system infrastructure and logical environmental considerations such as electromagnetic spectrum usage and interference. Finally, the time factor is defined by Conti and Raymond as including "key dates", "cultural perception" and synchronization issues [34]. In many operational environments, time is also an important interrelation element between kinetic and cyber operations.

The proposed paradigm can readily support the consideration of operational environmental factors as inputs to the security assessment process. It can also readily provide outputs that can be utilized as part of operational environment analysis processes. While not strictly within the paradigm, the Blackboard Architecture utilized for the implementation of the paradigm's tool can also support the development of networks for operational environment assessment.

The mechanisms for operational environment inputs and outputs are demonstrated in Figure 11. The figure shows how conventional facts and rules can be used to support this. Conventional facts are independent of containers and conventional rules interrelate between particular facts (instead of providing generic functionality that can be applied to any set of matching containers). As conventional rules can use specific property values within specific containers as their inputs and outputs, they can be used to interrelate operational environmental considerations with containers, their interrelating links and the generic rules that are used to define attacks.

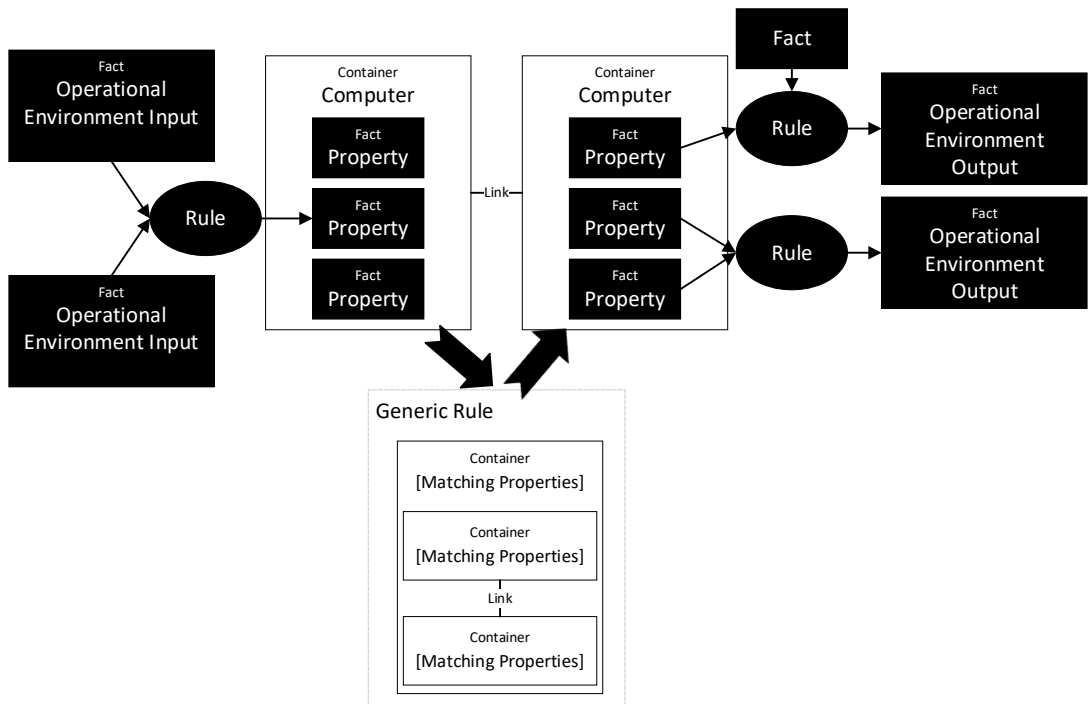

Figure 11. Example of using operational environment facts, conventional facts and conventional rules.

The types of impacts that the operational environment can have are numerous and varied. In the example of the scenario of hacking one computer from another, the operational environment could be integral to gaining the initial foothold on the left computer. For example, as shown in Figure 12, this could rely upon social engineering which is enabled by positive press coverage and public support (leading a user to provide the attackers with their credentials).

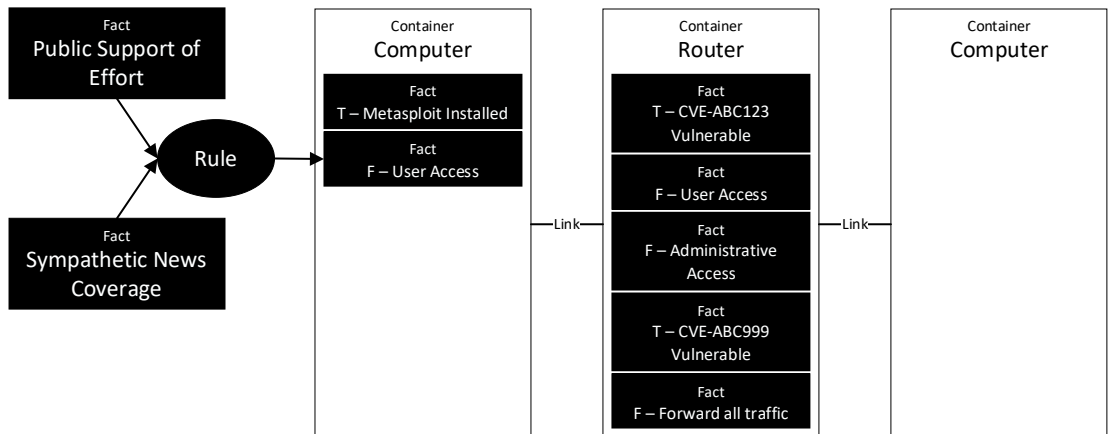

Figure 12. Example of the use of operational environment facts and rules.

This is implemented with two facts relating to the two operational environment factors (news coverage and public support), as shown in Figure 12. A rule is created to show how these two factors lead to gaining the password for the left computer, providing the access capability required to carry out the rest of the attack (as was previously shown).

### *4.7. War Games and Tabletop Exercises*

Another key area of use for the proposed paradigm is supporting war games and tabletop exercises. Both allow critical infrastructure operators, key decision-makers related to the critical infrastructure and those who depend on the critical infrastructure to plan for and play out scenarios where the infrastructure is targeted. War games and tabletop exercises may focus on cybersecurity threats specifically or may consider cybersecurity as part of a broader threat response exercise. In either case, the proposed paradigm can be used to provide insight regarding the status of the critical infrastructure before and during the tabletop exercise / war game scenario.

*4.7.1. Tabletop exercise*

A tabletop exercise is typically a lower-fidelity group discussion where a facilitator takes the group through a possible scenario and different actions and responses are discussed. In many cases, these focus on the decisions that leaders would make during such a scenario and their implications. The facilitator or other scenario staff may provide a narrative regarding the impact of the decisions that are made (or may ask the experts participating in the exercise what would happen based on the combination of their decisions and adversary actions and responses) to facilitate the scenario.

The proposed paradigm could support this type of exercise in two ways. The first way is by helping to answer questions regarding what could happen in response to a particular attacker or defender action. As opposed to making educated guesses, the action could be entered into the simulation system and the results provided. The second way is by providing a map, through visualization of the system, to aid participants' understanding of the system, generally, and the changes that attacker and defender actions make to it and its functionality throughout the exercise.

*4.7.2. War games*

A war game is typically a higher-fidelity exercise where a specific scenario is played out in detail. Some war games carry this to the extent of engaging actual vehicles and weapons in maneuvers. However, in the case of critical infrastructure systems, they need to operate continuously (and thus cannot be diverted from their actual use for war game participation) and typically won't have the ability to be moved (or have a spare / parallel system assigned) to a war game site. Thus, as opposed to using an operational system for a war game, the proposed paradigm and its implementing simulation system can be utilized.

At the most basic, the simulation system can be used to support a role-playing game-style war game where the simulation system provides the map for the game and indicates the results of different activities and actions. Each action is input into the system, in this case, and an up-to-date system status map can be provided to the simulation leader (who may provide only portions to participants, corresponding to the information that they have a mechanism to access).

A more advanced approach can combine simulation of some parts of the system (i.e., ones that cannot be easily duplicated or with which the participants are not going to interact) with emulation of other parts of the system. For example, the functional technology parts of the critical infrastructure system could be simulated, to provide a result to actions, while the IT systems are emulated, allowing the war game attackers and defenders to actually perform their respective roles using these systems.

In the case where combined simulation and emulation are used, the data collection tools (which are described in the next section) could be run frequently to update the system status. This would allow the simulation to import the current status of the emulated areas and extrapolate the impact to the simulated systems, based on the emulated areas' status.

**5. Data Collection Sources**

The previous section described how developed model networks can be utilized for security analysis purposes. Focus now turns to – and this section discusses – how model networks are created. Networks can be created from a variety of data sources, which are discussed in this section. The implications of the use of each type of data source are also discussed. These data sources include network diagrams, computing system information, network scans, probing of actual or clone systems, probing of twin and cousin systems, probing of similar systems and software fuzzing. Additional data can be added manually (including both system and failure implication data) or extrapolated. Each data source is now discussed.

*5.1. Network Diagrams*

Network diagrams are an important primary source of information for the construction of the model used by the proposed paradigm. The computing systems and networking hardware should be able to be identified from network diagrams and the diagrams should also provide information about the systems and networking hardware's interconnectivity. Thus, network diagrams should facilitate the creation of the containers that represent the computing systems and networking hardware and the links that represent the interconnectivity between them. Naming information from the network diagram can serve to provide names for the computing systems and networking hardware containers.

In some cases, network diagrams may be annotated with additional information beyond the existence, naming and interconnectivity of computing systems and networking hardware. For example, they may include annotations regarding the functionality of devices or IP addresses. This other information should also be captured, as appropriate to the particular simulation system.

*5.2. System Information*

While diagrams may provide an overview of the computing systems and networking hardware, additional details will be needed to fill in requisite details that are needed for effective simulation. Two types of system information are relevant. First, information about the specific hardware and software that comprise the various computing systems and networking hardware should be used to further enhance the fidelity of the simulation system. Second, configuration details regarding these systems can also add to this and greatly enhance the fidelity of the simulation through knowledge of the specific ways that the systems operate and any potential vulnerabilities that they may have present. Each will now be discussed.

5.2.1. Software and hardware details

The software and hardware details needed for the simulation will depend on its purpose and fidelity. Simulations that focus on the security of the IT systems will need details regarding hardware, to the extent that it can be matched with known vulnerability information. Information about the software installed on each system and its version (to the service patch or build level) will also be needed, again to the level needed to identify whether it is susceptible to any specific known vulnerabilities.

A simulation system that is designed to also analyze supply chain security will need further details about the individual hardware and software components of each system to facilitate building the nested containers that will represent this. Each component will need to be researched to identify if it has sub-components and to characterize the integrity of the supply chain up to and including that component (which will, in turn, be used to inform the analysis of the supply chain of systems and components that incorporate it).

5.2.2. Configuration details

The exploitability of vulnerabilities present in a system can be highly dependent on its configuration – and configuration issues can create vulnerabilities that are not present in the underlying software and hardware. Because of this, knowledge of the computing systems and network hardware configuration can enhance simulation fidelity significantly. Configuration details of networking equipment can also enhance the simulation's accuracy of simulating the performance of the network by providing greater details regarding how network traffic is processed (e.g., what traffic is forwarded across a router, is filtering being performed and what are filtering decisions based on).

Configuration details can be manually added to each relevant container and link in the simulation system's model. Alternately, automated tools can be used to probe the operational system or clone or twin systems to collect these details and add them to the simulation system model. Automated scanning is discussed in more detail in the subsequent subsections.

*5.3. Network Scans*

Network scans can provide valuable information about the operational system. This can include identifying deviations between design documentation (such as the network maps and configuration details described in the previous subsections) such as added, removed or differently configured computing systems or networking hardware. Three different types of network scans can provide value, in this regard. Conventional IP range and port scans, enumeration scanning and firewalking will be discussed.

*5.3.1. Conventional IP range and port scans*

Conventional IP address range and port scans, such as those performed using Nmap, can serve to identify deviations between the expected and actual network and computing system configuration. IP addresses for which a response is received and not expected may indicate the presence of new devices or configuration discrepancies. IP addresses from which a response is not received (and is expected) may indicate that systems have been removed, disabled or reconfigured. Ports which respond unexpectedly or don't respond as expected may indicate configuration changes or that systems have had their IP addresses reassigned. Differences in both port and IP responses from what is expected may also be informative about the configuration of network devices that the traffic is passing through, if applicable. Scanning from multiple vantage points can facilitate a better understanding of the impact of network devices on the scan results.

In addition to confirming expected results and identifying discrepancies and new devices, network scanning can also be used to build or fill in details for an initial model, if network maps and device information is not available.

*5.3.2. Enumeration scanning*

Tools that are designed to perform enumeration for cyberattacks and penetration tests can also be helpful for informing model development.  Like with the IP address range and port scans, these tools can identify unexpected services that may not be reflected in documentation. They can also identify services that may be configured differently than indicated in the documentation or which have been disabled. Enumeration goes beyond simply checking to see if ports are open (or not) to actually identifying what services are running on a port using banner grabbing and other techniques.

Like with IP range and port scanning, enumerations scanning can be used to fill in details (specifically, more granular information about what services are actually operating) for an initial model, if network maps and device information is not available.

*5.3.3. Firewalking of firewalls and routers*

Networking devices and – in particular – firewalls and routers can notably impact what attacks are possible against a critical infrastructure system.  By blocking or filtering traffic, for example, certain attacks can be prevented.  They may also limit adversaries' (and penetration testers and model builders) visibility of the network and the computing systems connected to it.

While details can be added to the model via manually assessing the configuration files for firewalls and routers, this approach may lead to errors as subtle differences between what is intended (and documented) and the interpretation of the actual configuration files by the devices may exist. Additionally, configuration files may have been changed since documentation was created or newly documented changes may not have been applied.  Thus, just with how IP range / port scanning and enumeration scanning can be used to validate the network model, firewalking can be used to validate the model's implementation of network device configurations.

Firewalking sends traffic through network devices to see what is forwarded and what is blocked. Traditional firewalking presumes that only a single interface is available to the scanner (and uses time to live values to attempt to ascertain where traffic is being dropped or filtered at); however, it can also be conducted – with higher accuracy – by connecting to multiple firewall or router interfaces.

Like with IP range, port and enumeration scanning, firewalking of firewalls and routers can also be used to build a model (of the applicable network devices), if configuration details are not available.

**5.4. Probing of operational or clone systems**

The previous subsections have discussed several types of scanning that can be utilized to characterize network devices and computing systems.  A key consideration when using these – or any other type of scanning or probing – is which version of systems to actually scan or probe.

Operational system components are the actual hardware that is used in the day-to-day operations of the critical infrastructure system.  While scanning and probing this hardware would provide the most accurate data, it also may trigger (or, via degradation to avoid it triggering, temporarily diminish the efficacy of) intrusion detection systems.  If scans or probing run the risk of impairing system performance, causing a denial of service or otherwise interfering with operations, this would be a large risk to undertake with operational system hardware.  While some scans could be conducted for validation, which are believed to be safe, it is possible that even believed-safe scans could cause issues (for example, a fast port scan might cause a denial of service if it overwhelms a computer or network device).

To reduce risk to operational systems, surrogates can be used.  Each type of surrogate (including the clones discussed in this subsection and twins and cousins discussed in the subsequent subsection) presents both benefits and drawbacks.

A clone is the closest surrogate to the actual system itself.  It is a direct copy of the operational system.  For example, a clone of a server may be made through hard drive duplication and then running the hard drive copy on another similar computer or as a virtual machine.  Network devices may be copied or, in the case of devices whose functionality is determined entirely by a firmware version and settings file, cloned be ensuring that the correct firmware is present on a virtual or physical copy and using the same settings file.

Clones will provide the most accurate information of any type of surrogate; however, they also involve creating copies of credential stores, logs and other files and specific settings which may present a risk to the operational system if they become known to adversaries.  Clones also require access to the actual operating system to make copies and copying may degrade system performance or even require taking systems offline for copying.

Clones, thus, mitigate some risks to the operational systems (as opposed to conducting scanning or testing directly on them); however, they also introduce new risks and fail to prevent other risks that are mitigated by working with twin and cousin systems.

*5.5. Probing of twin / cousin systems*

Like with probing the actual system or a clone, twin and cousin system can also be probed in the ways described previously (amongst others) and both networking hardware and computing systems of a twin or cousin system can be probed.  The difference between these is, similar to the difference between the actual and clone system, a trade-off between information accuracy and risk.

Twin systems are systems that are created from scratch (i.e., not from drive copies) using documentation and configuration data.  This can be actual configuration files or descriptions of how the system is supposed to be configured (including descriptions generated by scanning activities).  Twin systems, thus, should be highly similar – if documentation is accurate and any settings files used are current; however, they will lack some nuances of the system (such as non-documented changes, updates over time, log files and similar).  Probing and scanning twin systems, thus, provides information about the security of a system that is operating how the operational system is supposed to work.  This can certainly detect issues that may be relevant to the operational system; however, differences in interpretation of design documents and differences between the pristine twin system and potentially long-running operational system may make it so that certain operational system vulnerabilities are not present in the twin.

Cousin systems are even further removed from the operational system – and typically may be intentionally so.  The goal of a cousin system is to be of a similar type to the actual system so that principles and general categories of vulnerabilities and exploitation paths in the cousin system can be used to assess the operational system for similar potential issues.  Cousin systems, through abstraction, limit the utility of system information to attackers, should it fall into the wrong hands.  They also facilitate wider dissemination of results and analysis while mitigating the potential risk to the operational system from information sharing.  The differences between the cousin and the operational system make it so that operational system vulnerabilities and exploit paths may not be applicable to the cousin and vice-versa.  Thus, the role of the cousin will be to identify and demonstrate classes of vulnerabilities, vulnerability and exploit chain concepts and to demonstrate and assess prospective tools, paradigms, practices and protocols that could be applied to enhance the security of the operational system.  Cousin systems also facilitate the publication of security assessment research.

*5.6. Probing of similarly configured networks*

Even further removed from the operational system than clones, twins and cousins is the probing of similarly configured networks.  Of course, the utility of data from a completely different system to a model is quite limited.  However, it can be useful to fill in missing data and for extrapolation purposes.

When modeling an operational system that has operated for a long period of time, the scanning and probing of similar networks and comparison to their design documents may facilitate the identification of discrepancies that can be applied to a model.  These can be used to simulate changes over time and operational artifacts and effects when real world system change information isn't available.

Scanning and probing similarly configured networks can also be useful when attempting to create a cousin, as it can provide information that may be relevant to the design and which is not sourced from the operational system.  Thus, it adds nuance to the cousin without the risk that the nuance may disclose operational system details.

Finally, the scanning and probing of similarly configured networks (and potentially modeling them) can be used to identify issues and practices which may be applicable to the critical infrastructure system.  In this regard, the similarly configured network becomes a form of surrogate cousin system that can be assessed to find issues that may also be relevant to the primary critical infrastructure system.

*5.7. Fuzzing of software and hardware*

Fuzzing can be used to identify vulnerabilities in hardware and software.  This type of analysis can certainly be valuable to and included in a complete cybersecurity analysis of a critical infrastructure system.  Fuzzing can be used to assess all of a critical infrastructure system's hardware and software (using either production or non-production versions).  Alternately, it can be used to focus on areas of potential risk, such as hardware components and software that are not widely used and may not be thoroughly explored due to a lack of testing by others.

In addition to using fuzzing for vulnerability identification, fuzzing can also be used to collect data to inform the proposed paradigm's model creation.  By testing hardware and software systems with numerous prospective values for each possible setting (including, potentially, all valid values and numerous invalid ones), the operations of each hardware and software component can be characterized, allowing the prospective results of different types of attacker-produced configuration changes to be known. This also facilitates the assessment of settings files (and changes) based on the analysis of the settings file itself, based on the results of prior fuzzer-based testing.

## 5.8. Manually collected data

Manually collected data is a catch-all category for other types of information that might be added to a model created using the proposed paradigm. This data can be collected from interviews, published reports and numerous other data sources. While any sort of data that may be relevant to the cybersecurity assessment of a system might be added under this category, there are a few areas that may be of particular interest to critical infrastructure systems. The types of data that may be relevant will also depend heavily on how the model will be used.

### 5.8.1. Policies, procedures and response protocols

Complex cyberattacks may leverage knowledge of how an organization performs a given task or responds to an activity to trigger this task performance or response. Given this, a knowledge of organizational policies, procedures and response protocols that impact critical infrastructure system operations and access may enhance the utility of the model for assessing complex attack scenarios.

For example, an organization that, as a matter of policy, brings in outside contractors when a data breach is detected may be susceptible to an attack by individuals posing as these contractors. An adversary could create a data breach as a pretense to supply contractors to access parts of the system that they otherwise would not have access to. There are obviously a variety of permutations of this particular attack (including whether individuals are implanted at a consulting firm or a fake firm is created and markets its services to the critical infrastructure provider) that could be considered. Numerous other similar scenarios may be relevant to different critical infrastructure systems.

Data collected about policies, procedures and response protocols can be modeled using conventional rules and facts, as was described for modeling the operational environment. This allows the model to understand these factors and their impact on system component parameters. This data, as would be expected, can enhance the fidelity of the model and its efficacy for considering issues of this type.

### 5.8.2. Human operations data

Some areas of the operations of a critical infrastructure system may be performed by humans. This could include maintenance activities, activities as part of specific operating protocols or recurring actions that are performed on a regular schedule. Identifying activities that modify the system's state temporarily or on an ongoing basis can enhance system fidelity. This can facilitate the identification (and, thus, mitigation) of attacks that can only occur or attack paths that differ during periods of various human activities. Many human operations can be added to the model using conventional rules and facts, as was described for operational environment modeling.

### 5.8.3. Change-over-time information

Understanding how the system has and will change over time can help inform system models. Periods of change can present vulnerabilities as facilities are being modified, atypical staff members are present and systems are being configured.

Understanding the level of change that a system typically experiences can help analysts assess change-related risks and the frequency of model updates that are needed. An understanding of change processes can also be utilized to identify risks that may occur during system changes. This can include risks that may provide an opportunity for an adversary to gain a foothold that can continue to be used after the change process is complete and the vulnerability path used to gain the foothold has been closed.

*5.8.4. Discrepancy information*

Manual collection of discrepancy information can be used to help identify issues with the documentation that a model is based upon as well as issues that have been introduced due to mistakes during model creation. Collection can take the form of presenting knowledgeable individuals with system documentation or visualizations of the model and asking the to identify any areas of potential discrepancy or concern. These can be researched to ensure accuracy. Alternately, an individual (or group) could be tasked with manually verifying the system model on a node-by-node basis to ensure accuracy.

*5.8.5. Historical vulnerabilities and problems*

Historical vulnerabilities and problems can be informative regarding the types of issues that may occur in the future. Root cause analysis of these may indicate practices that lead to problematic configurations or other recurring issues. Thus, a knowledge of historical vulnerabilities and problems may inform the system model in a number of ways and help analysts identify procedural or systematic sources of potential vulnerability and system exploitation.

*5.8.6. Legacy and idled hardware*

In some cases, when a system is updated or replaced, components of previous system versions are left in place due to the expense or logistical issues with their removal. Thus, there may be components left that may be part of the system's attack surface that may not be documented in current system documentation. These may be accessible through legacy control systems or may even be inadvertently connected to phone lines or current networks. In some cases, these systems may have interconnectivity but use a protocol no longer in use by the primary control systems; however, they may still be able to be accessed using this protocol from control stations, if they are configured to do so.

Legacy and idled hardware may present a significant opportunity for attackers and could easily be left out of defensive calculus. Thus, manual collection regarding the presence of legacy and idled hardware can fill in these knowledge gaps.

5.8.7. Manual control systems and overrides

In some cases, critical infrastructure systems may have manual mechanism that can override or take actions similar to those available to the control system. These mechanisms may be utilized via a social engineering attack. Additionally, the presence of manual mechanism, if their state is not known to computing systems, may result in different results of actions than would otherwise be projected due to a difference between the actual and modeled system state. In some cases, manual mechanism could be in long-term use as a workaround for issues or due to a lack of knowledge and documentation.

Given this, gaining an understanding of the presence and use (both on an ongoing basis and in response planning) of manual control systems and overrides can enhance the fidelity and accuracy of the model of the critical infrastructure system and security assessment based on this model.

### *5.9. Implications of failure data*

In assessing the security of critical infrastructure systems as well as using the proposed paradigm's model for the other uses described in Section 4, knowledge of the implications of disablement or access to various system components can enhance the model. This information can facilitate the targeting of simulated attack runs (by focusing on the areas that attackers would get the most impact from). It can also aid in conducting tabletop exercises and war games, by providing information regarding the results of each prospective attack and defense action. Particularly if the functional technology of the critical infrastructure system is not being simulated, implications of failure data is crucial to understanding the vulnerabilities and potential attack chains relevant to (and most impactful to) the critical infrastructure system.

Notably, implications of failure data can be provided in terms of specific operational impacts, which would provide the greatest granularity of analysis. To protect this sensitive data, it can also be collected as severity-of-impact ratings. This limits follow-on analysis (such as analysis of the implications of two concurrent attacks to different areas of the critical infrastructure system); however, it also serves to prevent the potential exposure of an effective targets-and-impact list for the system. Notably, severity-of-impact ratings do not necessarily need to be ranked and can be provided in terms of categories (such as impacts to a particular capability of the critical infrastructure system) or any other grouping that is useful for reporting purposes (and which may further server to obfuscate the value of particular targets).

### *5.10. Extrapolation of possible configurations in the absence of data*

In some cases, relevant data may not be available from one or more of the above categories – or this data may be incomplete or out of date. Data collection in one or more areas may be cost prohibitive or access may be restricted to maintain the security of the operational system. In the case of simulating a planned change to a system, various details may not yet be known.

In any of these cases, different possible configurations should be developed to approximate the actual or planned system. When extrapolating, it will typically be desirable to develop multiple extrapolations that can be evaluated to see how sensitive the results (and key metric performance) are to the extrapolation particulars. Using multiple extrapolations facilitates the assessment of the non-extrapolated areas under multiple possible ways the extrapolated area may actually be (currently or as part of a possible change) implemented. If results are shown to be sensitive to the particulars of the extrapolation this can also indicate a need to fill in further details with actual (or, in the case of a change, planned) implementation settings, in order to facilitate high-accuracy analysis.

### 6. Paradigm Operations

Four different operating approaches to the use of the proposed paradigm have been developed. Each is now defined, described and evaluated. First, the use of the paradigm for the analysis of an operational system is presented. Next, digital operational twin operations are discussed. Then, digital functional cousin operations and, finally, the use of the paradigm for spot analysis are considered.

*6.1. Operational System*

In some ways, cybersecurity assessment of the operational system may be the simplest use of the proposed paradigm. Under this approach, we simply collect data from and about the system that is in production use. Most penetration testing, for non-critical infrastructure systems is conducted using this approach (in many cases, without a simulation and analysis system).

If a non-critical infrastructure system was being assessed, this process could be as simple as entering network and configuration data and using automated tools to collect other details about the system, software, hardware and network configuration.

What makes this approach difficult with critical infrastructure is the fact that every test or data collection mechanism that is going to be run on the operational system must be pre-tested to ensure that it cannot damage the system or impair its performance. This necessitates that another system be available for testing which mirrors the operational system. Thus, each test ends up needing to be run twice – once against a digital twin (or similar) for vetting and then once against the operational system.

Given that pre-testing doesn't completely remove the risk of damage to the operational system and that the expense of a twin system is also being incurred anyway, testing on the twin system is less time-consuming, less risky and, thus, more prudent. Notably, testing of a critical infrastructure system before it enters service (or during a shakeout period where it isn't supporting anything critical yet) wouldn't have these considerations and, during this phase of system operations, direct testing on the operational system would be simpler and present minimal risk.

*6.2. Digital Operational Twin*

The digital operational twin approach involves making a copy of the system based on using configuration files and documentation. The idea is that the twin should be a pristine system that is similarly configured to the operational system. While the two should operate very similarly, the pathway that the two systems take to their current status will be quite different. The operational system may have been operated for a significant period of time and upgraded multiple times. It may have legacy files, log files and other content that may present additional attack surface area.

The use of the digital twin, though, prevents risk to the operational system. It also facilitates the use of a system that can be changed and re-tested to see how possible changes and fixes impact system security. Digital twin systems may also provide benefit in terms of their capabilities to support troubleshooting and data collection.

With the digital twin, the process starts with twin creation and, then, validation. Once this is done, the process is as simple as entering network and configuration data into the modeling system and using automated tools to collect other details about the system, software, hardware and network configuration. These can be input into the simulation system to ascertain what vulnerabilities the system may have and what exploit paths may exist in it.

Changes and fixes can be applied to the twin system, tested and then potentially applied to the operational system, after their impact has been assessed and they have been further validated to not cause operational issues on the twin system.

*6.3. Digital Functional Cousin*

A digital functional cousin is a system that is designed to be similar to the operational system but not an exact duplication of it. The functional cousin is designed to facilitate the analysis of general types of vulnerabilities and exploitation paths without exposing the exact configuration of the operational system.

The process of creating a functional cousin is similar to that for making a digital twin. The functional cousin system is created from scratch. However, unlike the digital twin which uses settings files and documentation from the operational system, the functional cousin may use an intervening layer of abstraction or may be designed based on merging design features and other aspects of several related systems of the same type with the details of the operational system.

Once the functional cousin is developed, the process is (similar to the twin) as simple as entering network and configuration data into the modeling system and using automated tools to collect other details about the system, software, hardware and network configuration. These can be input into the simulation system to ascertain what vulnerabilities the system may have and what exploit paths may exist in it.

As these vulnerabilities and exploit paths will not necessarily be directly applicable to the operational system, they should be manually evaluated with an eye towards generalizing them into issues that numerous systems can be checked for. General solutions can also be developed and validated using the cousin. Since the cousin system isn't an exact duplicate, it would not be suitable for validating the security enhancements provided by or performance of planned software updates.

*6.4. Spot Analysis*

The last use scenario is spot analysis. The spot analysis approach makes use of the paradigm and simulation system to analyze a particular area of the operational (or twin / cousin) system independently from the rest of the system. To do this, the area of interest of the system is built out normally, using containers, links and generic rules. Other areas, though, are built out only in terms of their interactions with the area of interest. Inputs and outputs to the area of interest, for example, can be developed using conventional rules (though containers to serve as a boundary may be needed if the area of interest will interact with other areas through generic rule-implemented attacks).

The spot analysis approach allows precise control of all areas of the system not being evaluated at the moment to allow the analysis to focus on the targeted area. Additionally, sensitivity analysis can be conducted to see how different values in adjacent and interacting areas of the system impact the security analysis of the target area.

The principal benefit of this approach is that it facilitates rapid setup, as only the relevant part of the system needs to be implemented in the simulation system. The approach can also benefit from the use of the different data collection mechanisms and tools described in Section 5 to populate any areas of the system that need to be built. Care must be taken, though, to ensure that all interfaces between the spot that is targeted for evaluation and the rest of the system (and external environment) are implemented and set to relevant values. If areas of interaction are overlooked or inaccurate values are used the output of the analysis may be inaccurate.

## 7. Outputs and Implications

The proposed paradigm and its implementing simulation system's outputs will differ somewhat depending on how it is being used.  One of the most basic outputs from the system is a collection of attack paths that are possible through the system, based on the knowledge that the system presently has.  By default, the system will treat any attack pathway as possible unless there is something blocking it.  Thus, the system can be utilized to generate the complete collection of attack paths. While these can be manually reviewed for the purposes of validation of whether they can actually be exploited and to remediate them, a post-processing step can increase their value.

The collection of attack paths can be analyzed with frequency analysis techniques to find nodes and exploits that appear in the most paths to prioritize them for evaluation and possible remediation. Analysis can also focus on identifying nodes and exploits that provide access to certain critical areas and to identifying low-frequency-of-appearance nodes and exploits that may provide access to critical areas, thus driving a heightened need for further evaluation and remediation based on the capability that they may provide.

When particular remediation plans are developed, these can be implemented in the simulation system first to see what their impact is likely to be on system security.  For example, assessment could project whether the number of attack chains will be reduced (and to what extent) and whether new attack paths are created – and their targets, locations and other characteristics.

This type of analysis can be performed with actual operating system details or with twin or cousin systems.  As was discussed previously, the fidelity and direct utility of analysis conducted with twins and cousin systems is limited by the methods used to create them and the abstractions introduced in the cousin systems, in particular.

When performing supply chain analysis, similar outcomes can also be produced, including the identified attack paths through the layers of the supply chain.  Additionally, frequency analysis can also be used on the output attack paths to see which areas of the supply chain are involved the most.  This facilitates the identification of potentially problematic vendors, suppliers and parts for possible replacement or other remediation.  Of course, the other types of analysis that focus on identifying high frequency nodes and attacks for remediation activities can also be used during a supply chain assessment study.

As is illustrated by the foregoing, the proposed paradigm and simulation system can help to focus defensive activities on the areas of highest return.  In addition to this, the system can provide information to support various other types of analysis including supporting tabletop exercises and war games.

## 8. Conclusions and Future Work

This paper has introduced a paradigm for the penetration testing of systems that cannot be subjected to the risk of penetration testing.  Many critical infrastructure systems have this constraint, as their disablement or degradation may pose a risk to human live, safety and property.  While the risk of testing these systems is high, there is also an acute risk of not testing them and leaving them unprotected against adversaries' attacks, due to a lack of knowledge regarding their vulnerabilities.

To this end, the paper has discussed and demonstrated how the proposed paradigm can be implemented through a simulation system and how the data that is needed to preform these simulations can be obtained. It has also described how the paradigm and simulation system can be used to support the security evaluation of supply chain threats and how assessments of various levels of accuracy and fidelity can be performed, as part of a trade-off with the accuracy of the system data collected and input to the system.

The goal of the proposed paradigm and simulation system are to facilitate the identification of possible security problem areas of the system being evaluated in order to be able to direct additional security resources to these areas. Its efficacy for doing so has been discussed herein and several examples of its use have been presented.

A number of key areas of study remain for future work. Efforts are ongoing that use the system for security assessment purposes. Future work can, thus, focus on implementing updates to the paradigm and simulation system based on lessons learned from its use in this context. Additionally, other types of data may be available for some systems and developing ways to import and utilize these additional types of data to increase the fidelity of the system remains a needed area of prospective future study.

**Acknowledgements**

This work has been funded by the U.S. Missile Defense Agency (contract # HQ0860-22-C-6003). Thanks are given to Cameron Kolodjski, Cayden Schmidt, Jack Hance, Jonathan Rivard, Jordan Milbrath, Matthew Tassava and others working on this project who have helped refine the proposed paradigm through work on various systems and components that implement or support its simulation. Supporting technologies described in this work are described in more detail in papers under concurrent development.

**References**

1. Krishnan, S.; Wei, M. SCADA testbed for vulnerability assessments, penetration testing and incident forensics. *7th Int. Symp. Digit. Forensics Secur. ISDFS 2019* **2019**, doi:10.1109/ISDFS.2019.8757543.
2. Morris, T.; Srivastava, A.; Reaves, B.; Gao, W.; Pavurapu, K.; Reddi, R. A control system testbed to validate critical infrastructure protection concepts. *Int. J. Crit. Infrastruct. Prot.* **2011**, *4*, 88–103, doi:10.1016/J.IJCIP.2011.06.005.
3. Christiansson, H.; Luiijf, E. Creating a European SCADA Security Testbed. *Post-Proceedings First Annu. IFIP Work. Gr. 11.10 Int. Conf. Crit. Infrastruct. Prot.* **2007**.
4. Ficco, M.; Choraś, M.; Kozik, R. Simulation platform for cyber-security and vulnerability analysis of critical infrastructures. *J. Comput. Sci.* **2017**, *22*, 179–186, doi:10.1016/J.JOCS.2017.03.025.
5. Türpe, S.; Eichler, J. Testing production systems safely: Common precautions in penetration testing. *TAIC PART 2009 - Test. Acad. Ind. Conf. - Pract. Res. Tech.* **2009**, 205–209, doi:10.1109/TAICPART.2009.17.
6. Ralethe, S.G. Investigating Common SCADA Security Vulnerabilities Using Penetration Testing, University of the Witwatersrand, Johannesburg, 2014.
7. Rocha, F. Cybersecurity analysis of a SCADA system under current standards, client requisites, and penetration testing, Universidade do Porto, 2019.
8. Speicher, P. Simulated Penetration Testing and Mitigation Analysis, Universität des Saarlandes, 2022.

9. Knowles, W.; Baron, A.; McGarr, T. The simulated security assessment ecosystem: Does penetration testing need standardisation? *Comput. Secur.* **2016**, *62*, 296–316, doi:10.1016/J.COSE.2016.08.002.
10. Li, Y.; Yan, J.; Naili, M. Deep Reinforcement Learning for Penetration Testing of Cyber-Physical Attacks in the Smart Grid. *Proc. Int. Jt. Conf. Neural Networks* **2022**, *2022-July*, doi:10.1109/IJCNN55064.2022.9892584.
11. Greco, C.; Fortino, G.; Crispo, B.; Raymond Choo, K.-K. AI-enabled IoT penetration testing: state-of-the-art and research challenges. *Enterp. Inf. Syst.* **2022**, doi:10.1080/17517575.2022.2130014.
12. Brown, J.; Saha, T.; Jha, N.K. GRAVITAS: Graphical Reticulated Attack Vectors for Internet-of-Things Aggregate Security. *IEEE Trans. Emerg. Top. Comput.* **2022**, *10*, 1331–1348, doi:10.1109/TETC.2021.3082525.
13. Lee, S.; Kim, J.; Woo, S.; Yoon, C.; Scott-Hayward, S.; Yegneswaran, V.; Porras, P.; Shin, S. A comprehensive security assessment framework for software-defined networks. *Comput. Secur.* **2020**, *91*, 101720, doi:10.1016/J.COSE.2020.101720.
14. Wotawa, F. On the automation of security testing. *Proc. - 2016 Int. Conf. Softw. Secur. Assur. ICSSA 2016* **2017**, 11–16, doi:10.1109/ICSSA.2016.9.
15. Thompson, H.H. Why security testing is hard. *IEEE Secur. Priv.* **2003**, *1*, 83–86, doi:10.1109/MSECP.2003.1219078.
16. Hai, T. Artificial Intelligence in Cybersecurity. **2017**, *1*, 103.
17. Schwartz, J.; Kurniawati, H. Autonomous Penetration Testing using Reinforcement Learning. *arXiv Prepr.* **2019**.
18. Khan, S.; Parkinson, S. *Review into State of the Art of Vulnerability Assessment using Artificial Intelligence*; 2018; ISBN 9783319926247.
19. Patel, A.; Qassim, Q.; Wills, C. A survey of intrusion detection and prevention systems. *Inf. Manag. Comput. Secur.* **2010**, *18*, 277–290, doi:10.1108/09685221011079199.
20. Zhang, Q.; Zhou, C.; Xiong, N.; Qin, Y.; Li, X.; Huang, S. Multimodel-Based Incident Prediction and Risk Assessment in Dynamic Cybersecurity Protection for Industrial Control Systems. *IEEE Trans. Syst. Man, Cybern. Syst.* **2016**, *46*, 1429–1444, doi:10.1109/TSMC.2015.2503399.
21. Mohammad, S.M.; Surya, L. Security Automation in Information Technology. *Int. J. Creat. Res. Thoughts* **2018**, *6*.
22. Wright, L.; Davidson, S. How to tell the difference between a model and a digital twin. *Adv. Model. Simul. Eng. Sci.* **2020**, *7*, 1–13, doi:10.1186/S40323-020-00147-4/FIGURES/4.
23. Cimino, C.; Negri, E.; Fumagalli, L. Review of digital twin applications in manufacturing. *Comput. Ind.* **2019**, *113*, 103130, doi:10.1016/J.COMPIND.2019.103130.
24. Tao, F.; Zhang, H.; Liu, A.; Nee, A.Y.C. Digital Twin in Industry: State-of-the-Art. *IEEE Trans. Ind. Informatics* **2019**, *15*, 2405–2415, doi:10.1109/TII.2018.2873186.
25. Rasheed, A.; San, O.; Kvamsdal, T. Digital twin: Values, challenges and enablers from a modeling perspective. *IEEE Access* **2020**, *8*, 21980–22012, doi:10.1109/ACCESS.2020.2970143.
26. Fuller, A.; Fan, Z.; Day, C.; Barlow, C. Digital Twin: Enabling Technologies, Challenges and Open Research. *IEEE Access* **2020**, *8*, 108952–108971, doi:10.1109/ACCESS.2020.2998358.
27. Liu, M.; Fang, S.; Dong, H.; Xu, C. Review of digital twin about concepts, technologies, and industrial applications. *J. Manuf. Syst.* **2021**, *58*, 346–361, doi:10.1016/J.JMSY.2020.06.017.
28. Jones, D.; Snider, C.; Nassehi, A.; Yon, J.; Hicks, B. Characterising the Digital Twin: A systematic literature review. *CIRP J. Manuf. Sci. Technol.* **2020**, *29*, 36–52, doi:10.1016/J.CIRPJ.2020.02.002.
29. Hernandez, L.A.; Hernandez, S. Application Of Digital 3D Models On UrbanPlanning And Highway Design. *WIT Trans. Built Environ.* **1997**, *33*, 453, doi:10.2495/UT970361.
30. Tao, F.; Xiao, B.; Qi, Q.; Cheng, J.; Ji, P. Digital twin modeling. *J. Manuf. Syst.* **2022**, *64*, 372–389, doi:10.1016/J.JMSY.2022.06.015.

31. VanDerHorn, E.; Mahadevan, S. Digital Twin: Generalization, characterization and implementation. *Decis. Support Syst.* **2021**, *145*, 113524, doi:10.1016/J.DSS.2021.113524.
32. Straub, J. Modeling Attack, Defense and Threat Trees and the Cyber Kill Chain, ATTCK and STRIDE Frameworks as Blackboard Architecture Networks. *Proc. - 2020 IEEE Int. Conf. Smart Cloud, SmartCloud 2020* **2020**, 148–153, doi:10.1109/SMARTCLOUD49737.2020.00035.
33. Khan, M.S.; Siddiqui, S.; Ferens, K. A Cognitive and Concurrent Cyber Kill Chain Model. In *Computer and Network Security Essentials*; Springer International Publishing, 2018; pp. 585–602 ISBN 9783319584232.
34. Conti, G.; Raymond, D. *On Cyber*; Kopidion Press, 2017;